\documentclass[12pt]{article}
\usepackage{graphicx}

\newcommand\pubnumber{WSU--HEP--XXYY}
\newcommand\pubdate{\today}

\def\fsu{Department of Physics\\
Florida State University, Tallahassee, FL 32306, USA}
\def\support{\footnote{Work supported by NSF award no. 106489.}}

\usepackage{xspace}     
\newcommand{\pt}{\mbox{$p_T$}\xspace}
\newcommand{\raa}{\mbox{$R_{AA}$}\xspace}
\newcommand{\rdau}{\mbox{$R_{dAu}$}\xspace}
\newcommand{\rppb}{\mbox{$R_{pPb}$}\xspace}
\newcommand{\rcp}{\mbox{$R_{CP}$}\xspace}

\newcommand{\sqrtsn}{\mbox{$\sqrt{s_{_{NN}}}$}\xspace}
\newcommand{\jpsi}{\mbox{$J/\psi$}\xspace}
\newcommand{\psip}{\mbox{$\psi^{\prime}$}\xspace}

\newcommand{\pp}{\mbox{$p+p$}\xspace}
\newcommand{\auau}{\mbox{Au+Au}\xspace}
\newcommand{\HI}{\mbox{A+A}\xspace}

\newcommand{\uu}{\mbox{U+U}\xspace}

\newcommand{\dau}{\mbox{$d$+Au}\xspace}
\newcommand{\da}{\mbox{$d$+A}\xspace}
\newcommand{\ppb}{\mbox{$p$+Pb}\xspace}
\newcommand{\pa}{\mbox{$p$+A}\xspace}
\newcommand{\pbpb}{\mbox{Pb+Pb}\xspace}

\newcommand{\sigabs}{\mbox{$\sigma_{abs}$}\xspace}
\newcommand{\ups}{\mbox{$\Upsilon$}}

\def\bibliographystyle{\def\@bibstyle}%

\def\Title#1{\begin{center} {\Large #1 } \end{center}}
\def\Author#1{\begin{center}{ \sc #1} \end{center}}
\def\Address#1{\begin{center}{ \it #1} \end{center}}

\newcommand\pubblock{\rightline{\begin{tabular}{l} \pubnumber\\
         \pubdate  \end{tabular}}}
\newenvironment{Abstract}{\begin{quotation}  }{\end{quotation}}
\newenvironment{Presented}{\begin{quotation} \begin{center} 
             PRESENTED AT\end{center}\bigskip 
      \begin{center}\begin{large}}{\end{large}\end{center} \end{quotation}}

\def\beq{\begin{equation}}
\def\eeq#1{\label{#1}\end{equation}}
\def\eeqn{\end{equation}}

\def\beqa{\begin{eqnarray}}
\def\eeqa#1{\label{#1}\end{eqnarray}}
\def\eeqan{\end{eqnarray}}

\let\bar=\overbar

\def\Dslash{\not{\hbox{\kern-4pt $D$}}}
\def\dslash{\not{\hbox{\kern-2pt $\del$}}}

\def\msb{{\bar{\ssstyle M \kern -1pt S}}}

\begin{document}
\begin{titlepage}
\pubblock

\vfill
\Title{Experimental aspects of quarkonia production and suppression in cold and hot nuclear matter}
\vfill
\Author{Anthony D Frawley\support}
\Address{\fsu}
\vfill
\begin{Abstract}
When heavy Quarkonia are formed in collisions between between nuclei, their production cross section is modified
relative to that in \pp collisions. The physical effects that cause this modification fall into two categories. Hot matter effects
are due to the large energy density generated in the nuclear collision, which disrupts the formation of the 
quarkonium state. Cold nuclear matter effects are due to the fact that the quarkonium state is created in a nuclear target.
I will review experimental aspects of quarkonia production due to both hot and cold matter effects.
\end{Abstract}
\vfill
\begin{Presented}
The 7th International Workshop on Charm Physics (CHARM 2015)\\
Detroit, MI, 18-22 May, 2015
\end{Presented}
\vfill
\end{titlepage}
\def\thefootnote{\fnsymbol{footnote}}
\setcounter{footnote}{0}
%

\section{Introduction}

\begin{figure}[htb]
\centering
\includegraphics[height=3in]{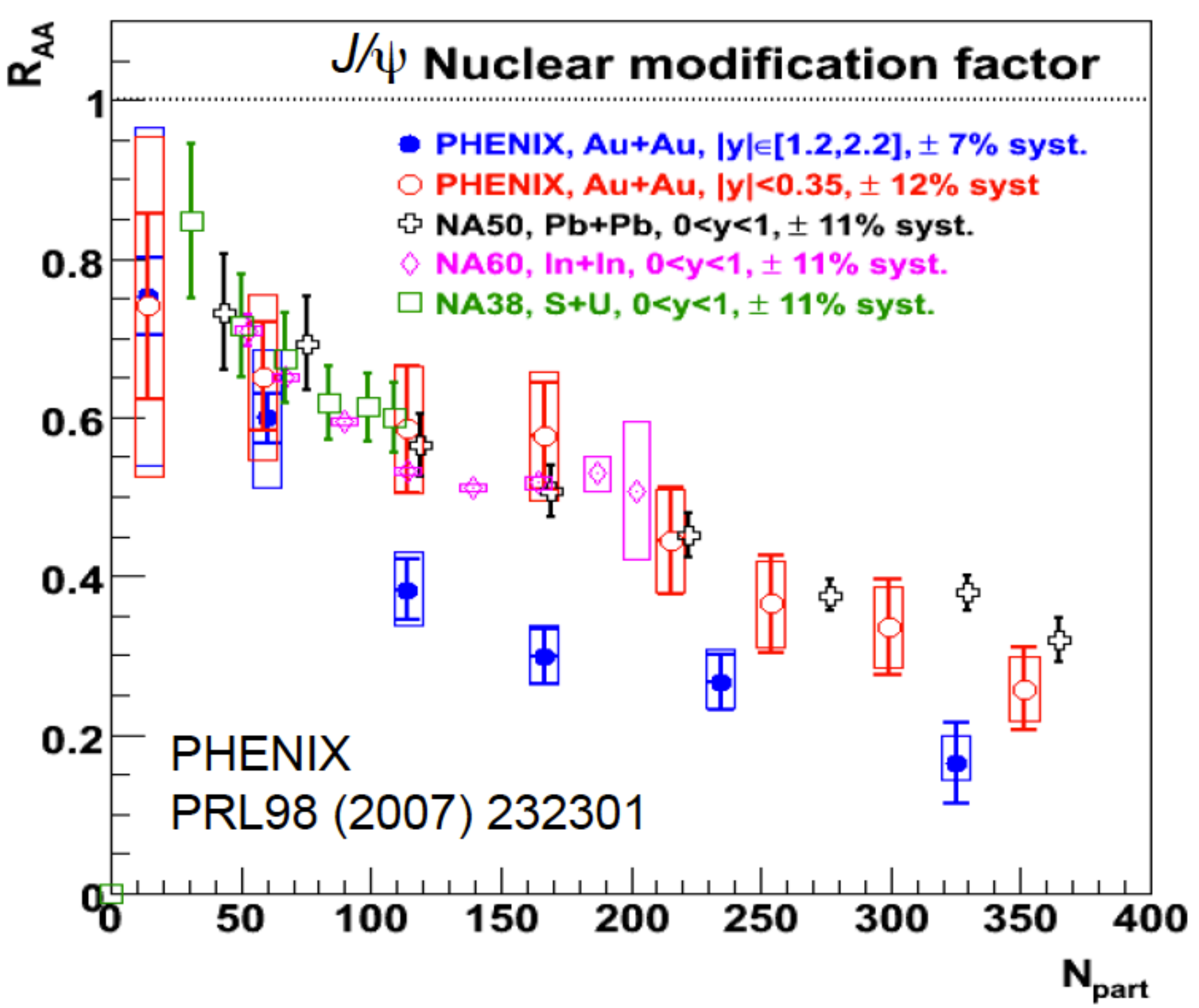}
\caption{The nuclear modification factor versus collision centrality for collisions of heavy nuclei at energies of \sqrtsn = 17.3 (In+In, Pb+Pb), 19.4 GeV (S+U), 
and 200 GeV (\auau).}
\label{fig:raa-uncorrected}
\end{figure}

Interest in studying  quarkonia production in high energy heavy ion collisions was motivated decades ago by the prediction that \jpsi formation would be
suppressed by color screening effects in the Quark Gluon Plasma (QGP)~\cite{Matsui:1986dk}. But experience has shown that relating \jpsi suppression to 
the energy densities 
of the hot matter formed in heavy ion collisions is complicated by the presence of competing effects that also modify \jpsi production. 
The importance of this is illustrated in Figure~\ref{fig:raa-uncorrected} where the 
nuclear modification factor, \raa, is plotted versus collision centrality for collisions of heavy nuclei at \sqrtsn = 17.3 to 200 GeV. There is no obvious dependence on 
the energy density produced in the collision. The measured nuclear modification factors
are similar at mid rapidity at the two widely different collision energies, but at 200 GeV show more suppression at forward rapidity,  where the energy density 
is slightly smaller than at mid rapidity. Interpreting the data requires a detailed understanding of the mechanisms that compete with color screening.

In general, quarkonia production is modified 
by both cold nuclear matter (CNM) effects and hot matter effects~\cite{Brambilla:2010cs} 
CNM effects modify the yield or kinematic distributions of \jpsi produced in a nuclear target in the absence of a QGP. They include modification of parton densities
in a nucleus~\cite{Hirai:2007sx, Eskola:2009uj, Kovarik:2010uv, deFlorian:2011fp, Helenius:2012wd}, breakup of the charmonium or bottomonium or its 
precursor $Q\bar{Q}$ state in the 
nuclear target due to collisions with nucleons~\cite{Lourenco:2008sk, Arnaldi:2010ky, McGlinchey:2012bp}, transverse momentum broadening as the $Q\bar{Q}$ 
traverses the 
cold nucleus, and initial state parton energy loss~\cite{Arleo:2012hn}.  CNM effects are expected to be strongly
dependent on collision system, collision energy, rapidity and collision centrality. 

In hot matter, there is not only suppression of quarkonia yields by color screening effects that reduce the binding energy of the quarkonium state, but also enhancement of
quarkonia yields due to coalescence of $Q\bar{Q}$ pairs that are initially unbound, but which become bound due to interactions with the medium (see for example~\cite{Zhao:2010nk}). In particular, at LHC collision energies the yield of \jpsi from coalescence of a $c$ and $\bar{c}$ from different hard processes becomes dominant.

When discussing modification of quarkonia yields in nucleir collisions, the hierarchy of relevant time scales is very important. The nuclear crossing time for 
Lorentz contracted nuclei at RHIC(LHC) energies is $\sim 0.3$(0.001)~fm/$c$. The formation time of a \jpsi meson in its rest frame is $\sim 0.3$~fm/$c$. The 
QGP thermalization time is $\sim 0.3-0.6$~fm/$c$. The QGP lifetime in central collisions at RHIC energy is $\sim 5-7$~fm/$c$, longer at LHC energies. 
The \jpsi lifetime is $\sim 2000$ fm/$c$. Thus the creation of quarkonia
and their modification in the hot medium occur on different time scales. They are often taken as factorable, although there are some theoretical models in which
they can not be factorized~\cite{Kopeliovich:2010nw}. 

\section{proton-Nucleus collisions as a reference for CNM effects}

CNM effects are observed in either \pa or \da collisions, which allow us to study the modification of quarkonia yields when they are produced in a nuclear target. 
It should be noted here that \pa collisions involving heavy targets appear to produce a small hot spot containing a large energy density, as evidenced by the observation 
of collective flow effects~\cite{Khachatryan:2015waa}. However when studying hard probes (jets or heavy quarkonia), \pa collisions differ from \HI collisions. 
In \pa collisions the hot matter and the 
hard probes  are both produced in a transverse area defined by the radius of the projectile proton. In \HI collisions however a high energy density is produced 
throughout the collision volume, and the hard probe is typically immersed in that extended volume of evolving hot matter for several fm/$c$ or more. We will return later
to the question of whether the small hot spot produced in \pa collisions significantly modifies the yield of hard probes.

There have been several studies of \pa or \da data in which \pt integrated \jpsi \raa values were corrected for shadowing effects using a shadowing
parameterization, and then the data were fitted to extract an effective absorption cross section, \sigabs, where \sigabs accounts for all effects 
that modify \jpsi production aside from shadowing. If it is assumed that CNM effects can be factorized from hot matter effects, then heavy ion collision data 
can be corrected for CNM effects using this parameterization of the \pa data.

\begin{figure}[htb]
\centering
\includegraphics[height=2.8in]{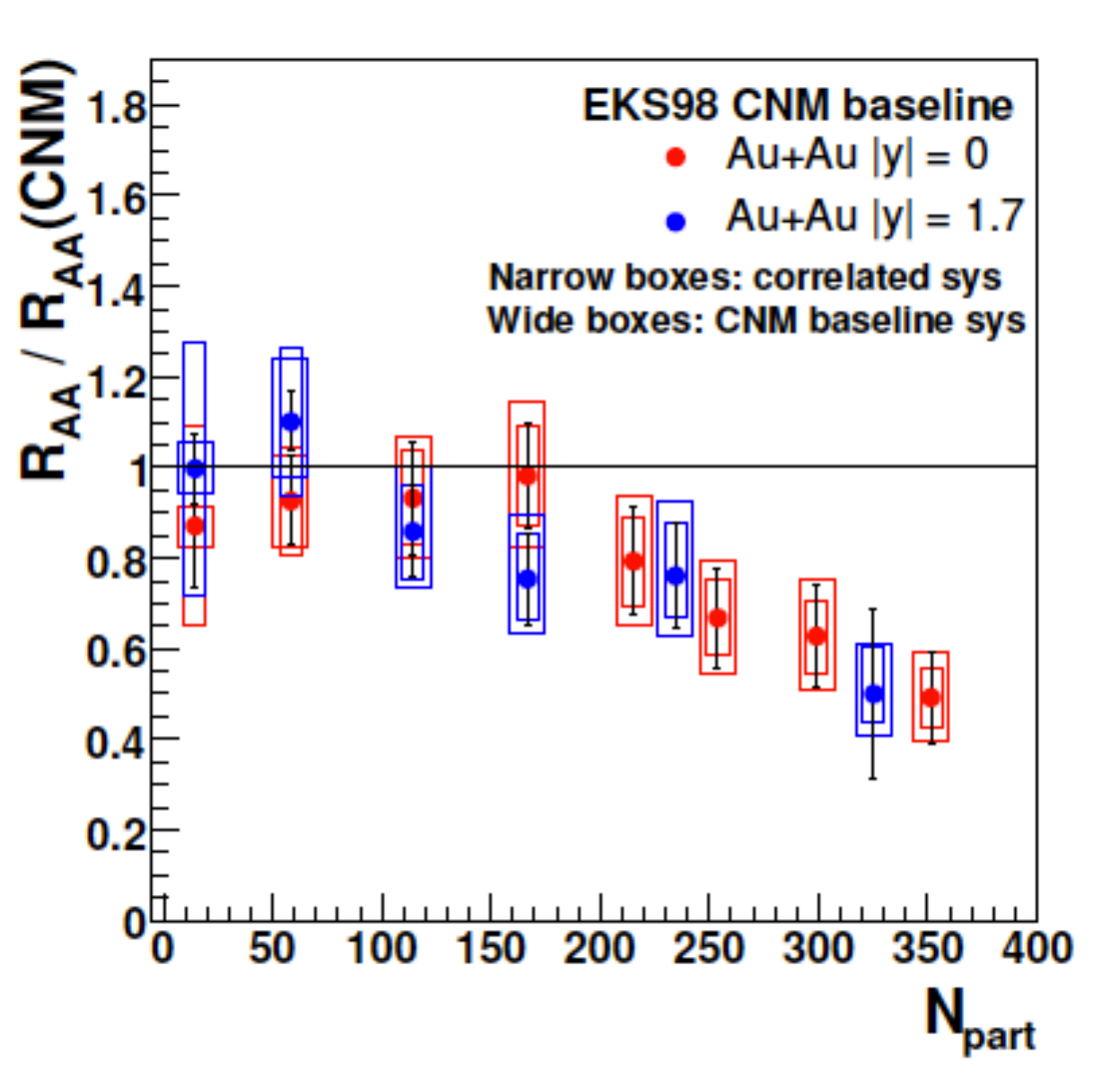}
\includegraphics[height=2.8in]{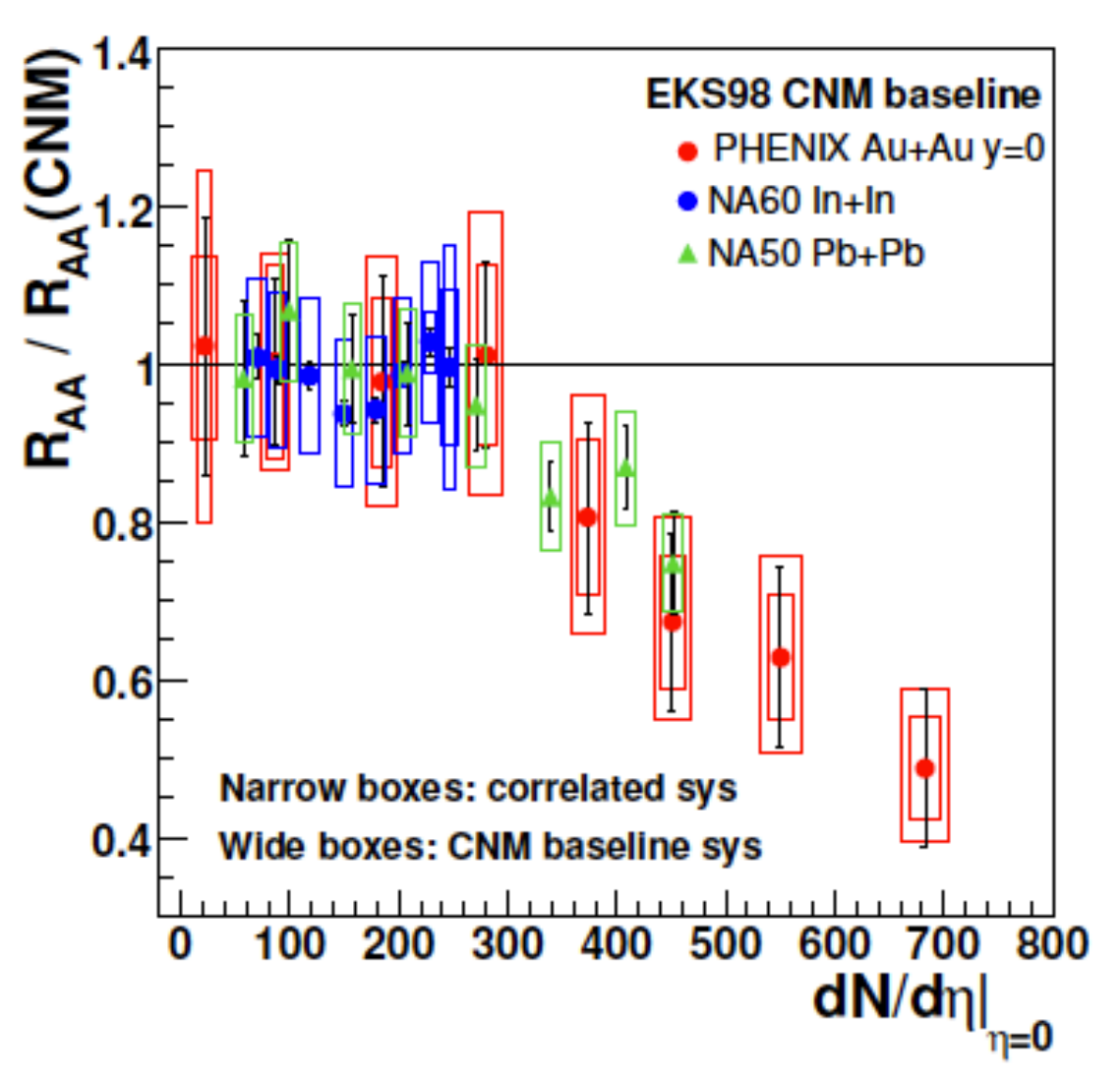}
\caption{Left: The measured \raa for \jpsi production at 200 GeV at $y=0$ and $y=1.7$ divided by the estimated \raa due to CNM effects.
Right: Comparison of the measured \raa divided by the estimated \raa due to CNM effects for \jpsi production in \pbpb collisions at sqrtsn = 17.3 
GeV at $y=0.5$ and in \auau collisions at \sqrtsn = 200 GeV at $y=0$}
\label{fig:raa-cnm-corrected}
\end{figure}

In one such study (see section 5 of~\cite{Brambilla:2010cs}) \dau \rcp data from PHENIX at \sqrtsn = 200 GeV were corrected for shadowing effects 
using the EPS09 parameterization and
values of \sigabs extracted at rapidities of $ -2.2<y<-1.2y$, $y < 0.35$ and $1.2<y<2.2$. The parametrization was then used in a Glauber model to estimate
the contribution to the \auau \raa from CNM effects at $y < 0.35$ and $1.2<|y|<2.2$, and the measured \auau \raa was divided by the CNM \raa. The result is
shown in Figure~\ref{fig:raa-cnm-corrected} (left). Comparison with Fig.~\ref{fig:raa-uncorrected}  shows that the correction for CNM effects eliminates the 
large difference in suppression for the \auau data at mid and forward rapidity.

In a similar study of \ppb data from the NA60 experiment at \sqrtsn = 17.3 GeV, the \raa was corrected for shadowing effects using the EKS98 parameterization,
and then the value of \sigabs extracted~\cite{Arnaldi:2010ky}. This was used to estimate the contribution to the \raa in \pbpb collisions so that the measured \pbpb \raa
could be corrected for CNM effects. In Figure~\ref{fig:raa-cnm-corrected} (right) the resulting CNM corrected results at \sqrtsn = 17.3 GeV and $y = 0.5$ are compared with 
the CNM 
corrected mid rapidity data  for \auau at \sqrtsn = 200 GeV. The data are presented as a function of $dN/d\eta$, which is used as a proxy for energy density. 
It is found that when the \HI data are corrected for CNM effects the remaining nuclear modification seems to scale with energy density, and is about 50\% for
central collisions at RHIC energy. 

\section{Nucleus-Nucleus collisions and coalescence}

As mentioned earlier, in addition to suppression of the quarkonia due to color screening, there is another effect that can occur due to QGP formation - coalescence.
There are two coalescence scenarios. 

In the first, a $Q$ and $\bar{Q}$ that were produced in the same hard process, and are nearly bound,
can become bound through interactions with the medium. This first scenario occurs even if only one heavy quark pair is produced in a 
collision, and is present to some degree at all collision energies. It is sometimes referred to as "regeneration".

In the second scenario, a $Q$ and $\bar{Q}$ that were produced in different hard processes can thermalize in the medium and 
combine statistically at hadronization. This second scenario becomes important only if a single collision produces many heavy quark
pairs. At LHC energies we expect that $\sim 100$ $c\bar{c}$ pairs are created in a central \pbpb collision.

\begin{figure}[htb]
\centering
\includegraphics[width=3in]{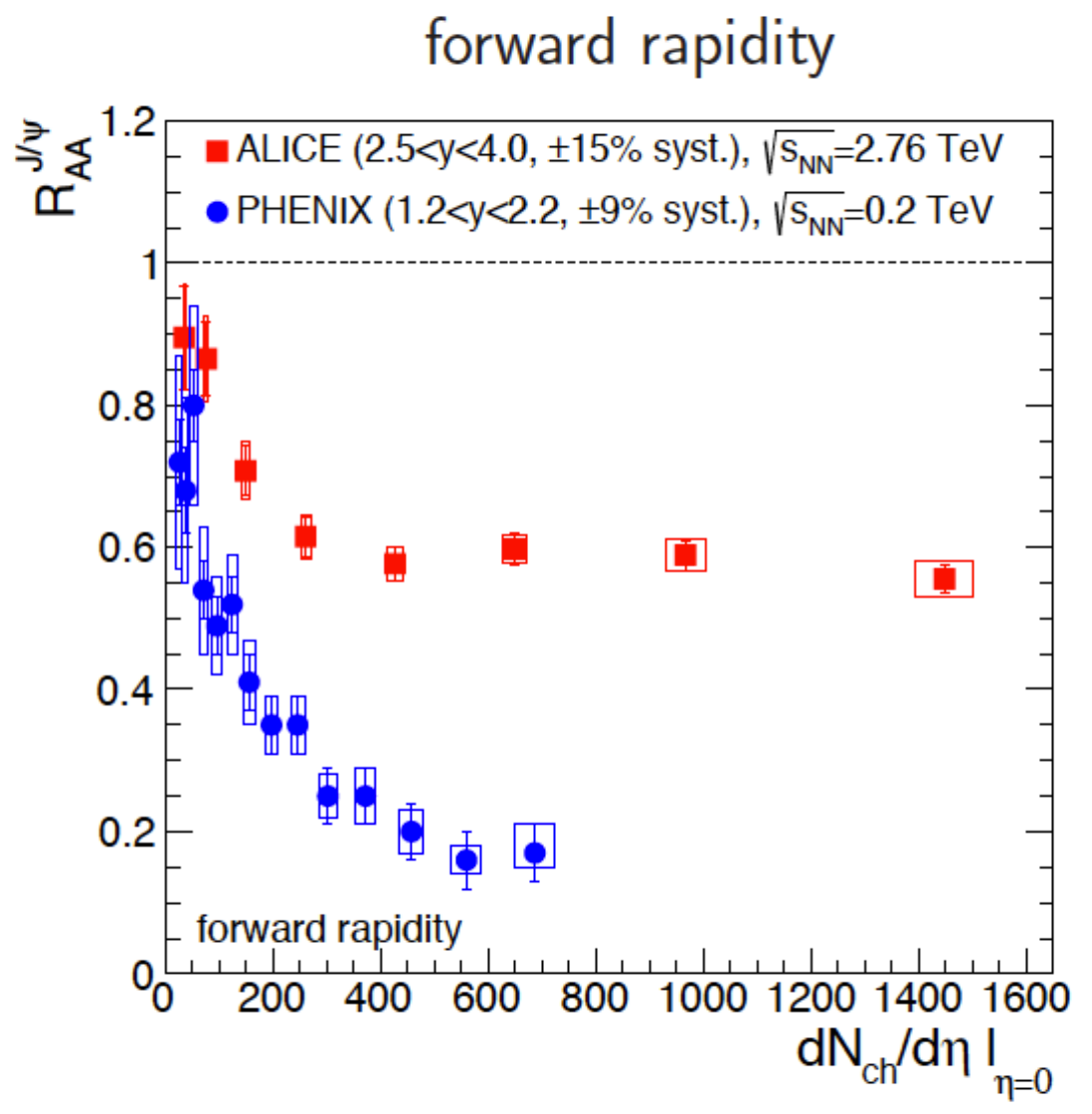}
\caption{Comparison of \jpsi \raa data at \sqrtsn = 2.76 TeV by ALICE with \jpsi \raa data by PHENIX at 200 GeV. }
\label{fig:alice-phenix-jpsi-raa}
\end{figure}

Figure~\ref{fig:alice-phenix-jpsi-raa} shows a comparison of measured \raa values for \sqrtsn = 2.76 TeV ALICE \pbpb results and 
\sqrtsn =  200 GeV PHENIX \auau results. The data, measured at forward rapidity in both cases, are plotted versus $dN_{ch}/d\eta$, 
which is used as a proxy for energy density. The suppression for the lower energy data is far stronger 
than for the higher energy data, suggesting that coalescence is now dominant. Indeed, the increase 
in \jpsi yield at the higher energy is concentrated at lower transverse momentum, as predicted by models that include coalescence 
at LHC energies~\cite{Adam:2015jsa}. 

\begin{figure}[htb]
\centering
\includegraphics[height=3in]{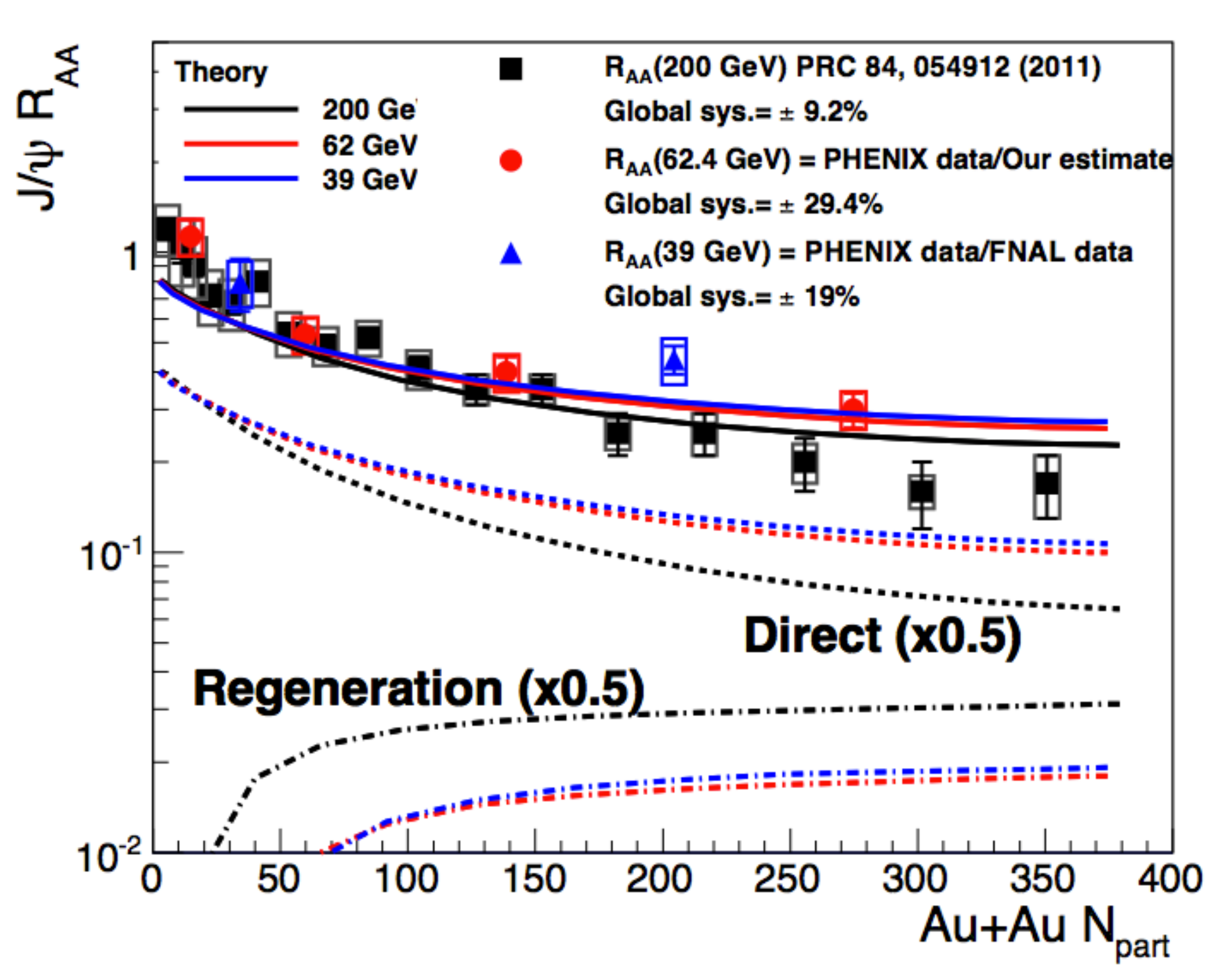}
\caption{The \raa for \auau collisions at \sqrtsn = 39, 62 and 200 GeV from PHENIX, compared with calculations that include both suppression due
to color screening and coalescence.}
\label{fig:energy-scan}
\end{figure}

Of the \HI data available so far, the strongest suppression occurs at RHIC energy, presumably because at the higher LHC energies the 
gain in \jpsi yield overcomes the loss due to stronger color screening effects. It is interesting to look at the dependence of the 
suppression for \auau on collision energy. Figure~\ref{fig:energy-scan} shows \raa vs centrality for \auau collisions 
at \sqrtsn = 39, 62 and 200 GeV~\cite{Adare:2012wf}. The data are compared with theoretical calculations that include coalescence~\cite{Zhao:2010nk}. The 
measured \raa is found to be similar at all three energies, and in the calculations that similarity is expected as a result of the 
stronger suppression due to color screening being balanced by an increased coalescence component as the energy increases.

In addition to the \auau data, there are preliminary \uu \jpsi data at \sqrtsn = 193 GeV from PHENIX at forward rapidity. They show slightly weaker suppression 
for central collisions than is seen in \auau data. This is consistent with a picture in which the increased suppression from the higher energy density in the \uu
collision is more than balanced by the increased charm production from the larger number of nucleon-nucleon collisions in \uu~\cite{Kikola:2011zz}.

\section{CNM effects in proton-Nucleus collisions}

\begin{figure}[htb]
\centering
\includegraphics[height=3in]{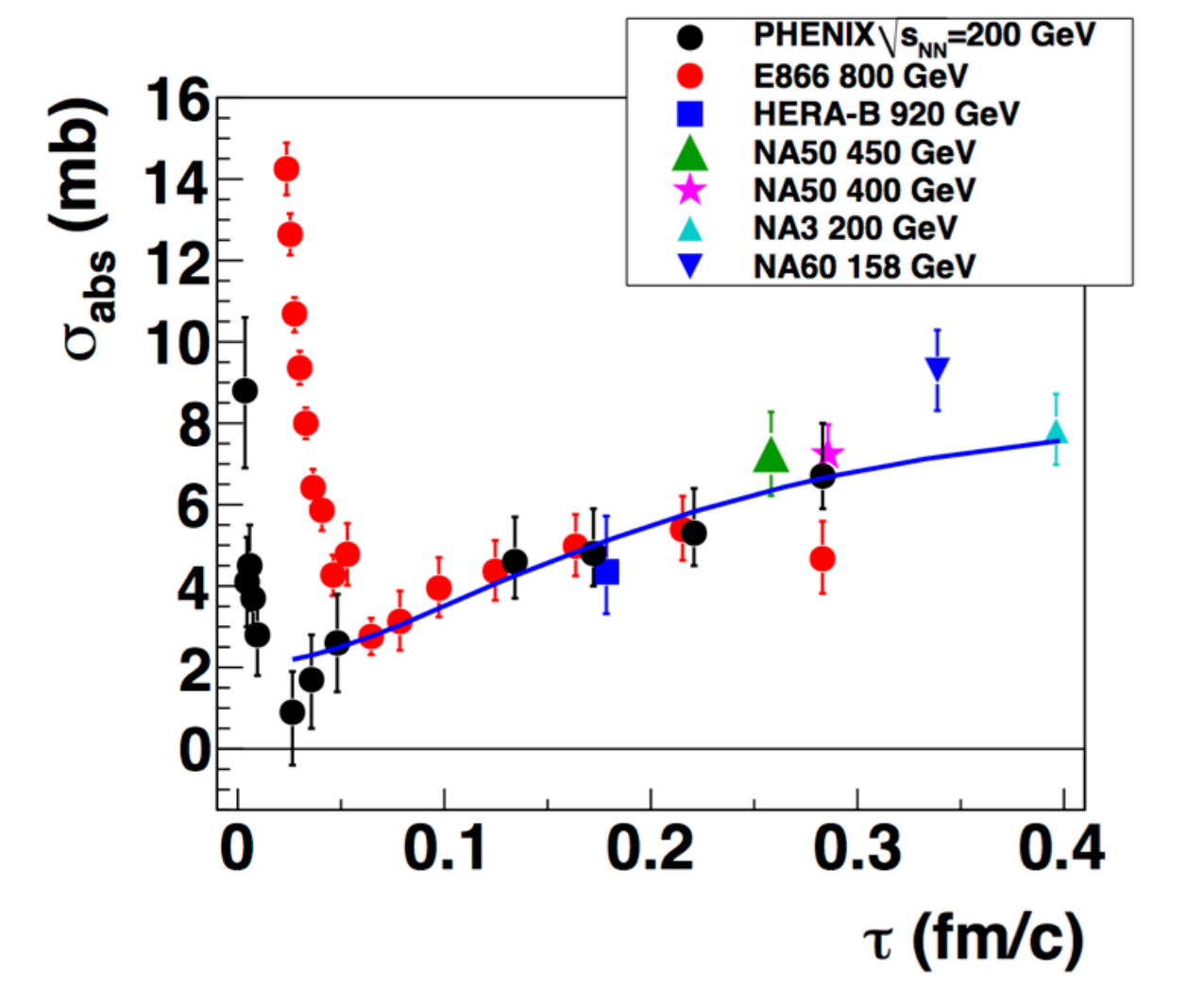}
\caption{Effective absorption cross sections extracted from shadowing corrected \pa or \da data at a range of collision energies from
\sqrtsn = 17.3 to 200 GeV, plotted versus proper time spent in the target nucleus by the evolving charmonium state. The curve is discussed
in the text.}
\label{fig:sigabs-tau}
\end{figure}

There was a comprehensive study of fixed target \jpsi \pa data from five fixed target experiments~\cite{Lourenco:2008sk}  in which the measured \raa values for centrality 
integrated data were corrected for shadowing effects using the EKS98 parameterization, and then fitted with an effective absorption cross section. Additionally 
there is a
value of \sigabs for the \jpsi from EKS98 corrected \ppb data at \sqrtsn = 17.3 GeV from~\cite{Arnaldi:2010ky}. A similar study of \jpsi \raa values from \sqrtsn = 200 
GeV \dau collisions at
nine rapidities, this time corrected for shadowing using the EPS09 parameterization, was made in \cite{McGlinchey:2012bp}. The \sigabs values extracted 
from all seven data sets
were plotted versus the nuclear crossing time - the proper time in the rest frame of the evolving charmonium state during which it is exposed to collisions with nuclei. 
The results from~\cite{McGlinchey:2012bp} are presented in Figure~\ref{fig:sigabs-tau}. 

For nuclear crossing times of $\tau > 0.05~fm/c$ the data at different energies seem to scale well with $\tau$.  The data at $\tau > 0.05 fm/c$ were 
fitted~\cite{McGlinchey:2012bp} with a model of an expanding color neutral meson passing through the target~\cite{Arleo:1999af}, showing that the 
data in this $\tau$ range are 
consistent with breakup of charmonia. For shorter nuclear crossing times no such scaling is observed. This is expected, since those nuclear crossing 
times are shorter than the $c\bar{c}$ formation time, suggesting that for short nuclear crossing times the effective absorption cross section reflects physical
processes other than breakup of the charmonia, such as energy loss in cold nuclear matter.

The observation of elliptic flow in \pa collisions at LHC energies and \da collisions at RHIC energies (see for example~\cite{Khachatryan:2015waa}) 
suggests that a small fireball is produced in those collisions. This raises the question of whether quarkonia production in \pa collisions is modified by 
final state effects. 

For the \jpsi, the \pa data from ALICE at the LHC seem to be consistent with models of CNM effects, including coherent energy loss at forward
rapidity~\cite{Adam:2015jsa}, although there are significant experimental and theoretical uncertainties. At RHIC energy, with
similar caveats, there is no strong evidence that more than CNM effects are needed to describe the PHENIX \dau \jpsi data~\cite{Adare:2012qf}.
It is worth noting also that the scaling observed with
$\tau$ in Fig.~\ref{fig:sigabs-tau} for $\tau > 0.5~fm/c$ would be broken if there were strong final state effects in \dau collisions at 200 GeV, but it seems to
hold within the uncertainties on the extracted \sigabs values. In all cases, however, because feed down from the \psip accounts for only about 10\% of the inclusive \jpsi
yield, it is still possible that the very weakly bound \psip is significantly modified by final state effects. 

\begin{figure}[htb]
\centering
\includegraphics[height=3in]{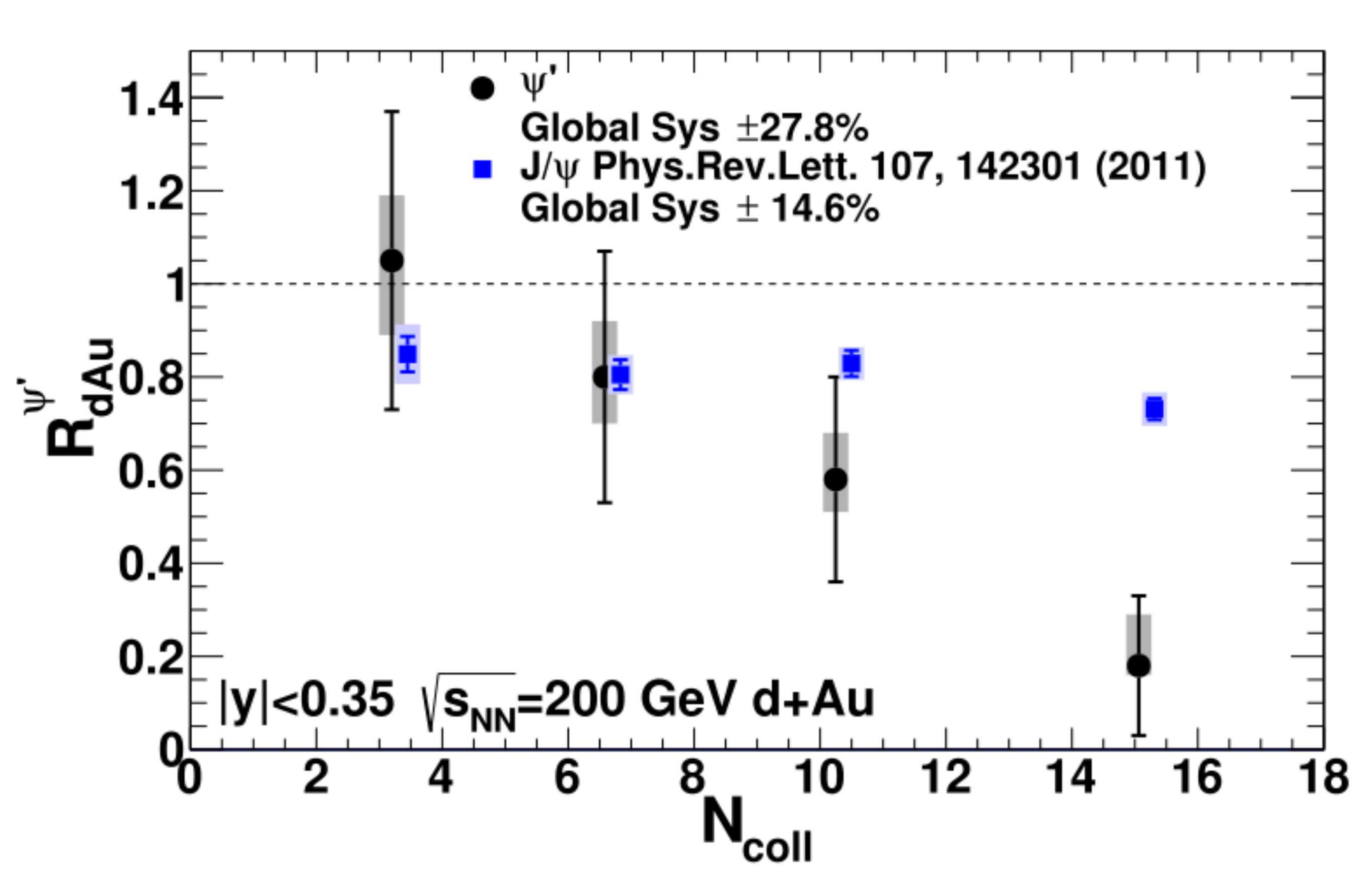}
\caption{The \rdau for \psip production in \dau collisions at \sqrtsn = 200 GeV measured at mid rapidity by PHENIX. The \rdau for \jpsi production is shown for comparison.}
\label{fig:psiprime-rdau}
\end{figure}

Unexpectedly strong suppression of \psip yields in central \dau collisions was observed at \sqrtsn = 200 GeV at mid rapidity by PHENIX~\cite{Adare:2013ezl}, as
shown in Figure~\ref{fig:psiprime-rdau}.
A very similar strong suppression of the \psip yield is observed in central \ppb collisions at \sqrtsn = 2.76 TeV at both forward and backward rapidity in 
ALICE preliminary data~\cite{Arnaldi:2014kta}. The strong differential suppression of the \psip relative to the \jpsi can not be due to a larger nucleon breakup cross section
for the larger, more weakly bound \psip because the nuclear crossing time is too short at all of the collision energies and rapidities of the data. Nor is it 
expected from any other known CNM process. It seems likely that the strong \psip suppression is caused by final state effects. Destruction of the \psip by
co-moving particles in the final state has been proposed as an explanation~\cite{Ferreiro:2014bia}.

\section{Upsilons}

The comparison of charmonium measurements at RHIC and LHC energies has revealed some very interesting physics. However it has not provided a 
direct comparison of color screening effects at RHIC and LHC temperatures because the dominant mechanism for charmonium production at LHC
energies, charm coalescence, is different from that at RHIC energies, where color screening dominates.

The study of Upsilon production in \HI collisions offers several advantages over charmonium. First, the \ups(1S), \ups(2S) and \ups(3S) states all have
significant branches to dileptons, and can thus be observed in the same experiment. The three states span a broad range of binding energies and sizes,
and have similar shadowing effects. Importantly, the \ups~states will not have large coalescence contributions at RHIC or LHC because the bottom production rate 
in central collisions at 
LHC energies is similar to the charm pair production rate at RHIC. On the down side, the \ups~production cross sections are small, and the difference in mass 
between the states is also small. Therefore mass resolved \ups~measurements  require large luminosity, large detector acceptance, and excellent momentum resolution.

CMS has measured the ratios of \ups(2S) and \ups(3S) to \ups(1S) in \pp and \ppb collisions~\cite{Chatrchyan:2013nza}. In \ppb collisions they observe that the \ups(2S) 
and \ups(3S) are differentially
suppressed relative to the \ups(1S) by about 17\% and 30\% respectively. The differential suppression of the excited states is found to
be stronger for events with larger particle production, suggesting that the larger number of particles in the final state has a stronger effect on the more weakly 
bound states.

\begin{figure}[htb]
\centering
\includegraphics[height=3in]{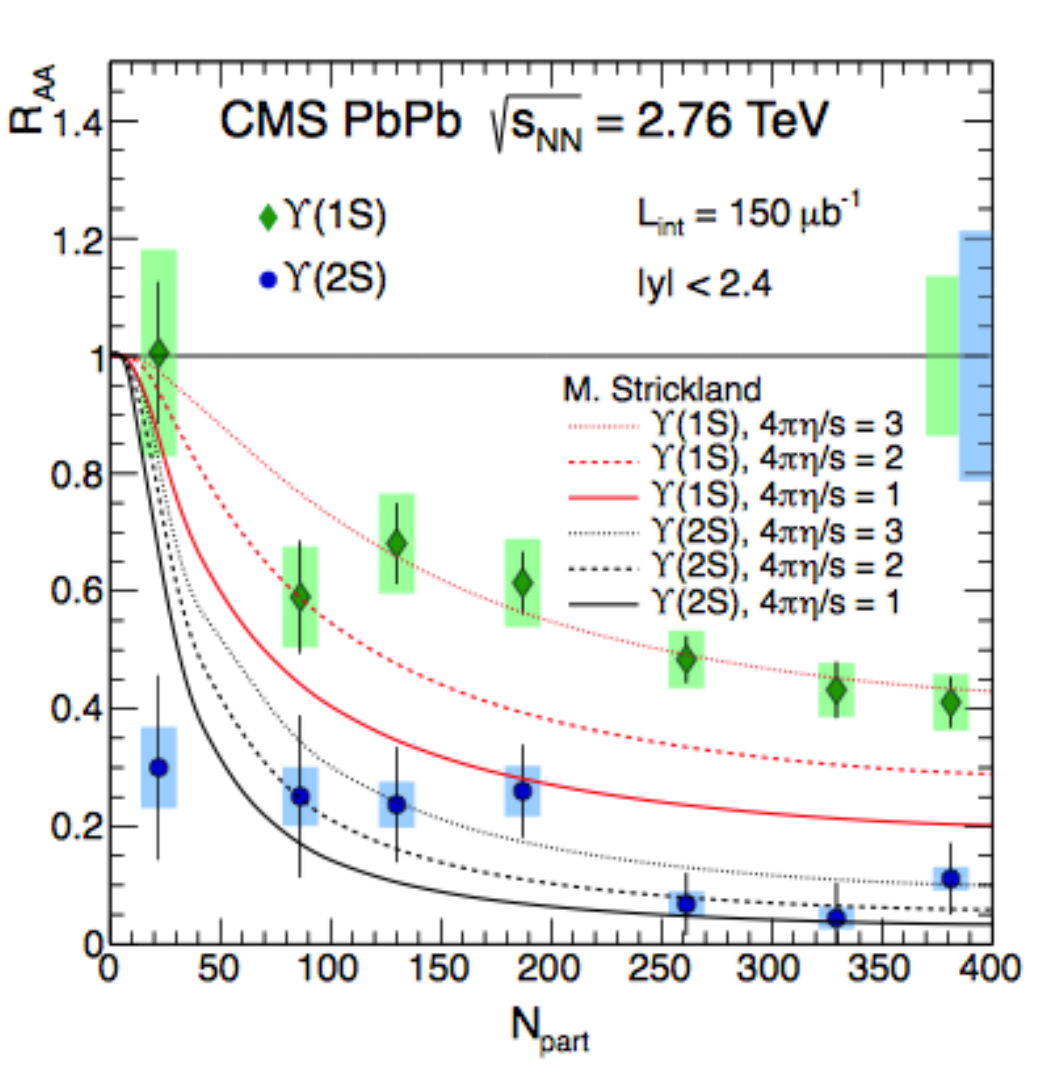}
\caption{The \rppb for \ups~production at \sqrtsn = 2.76 TeV measured at mid rapidity by CMS. The theory curves are from~\cite{Strickland:2011aa}.}
\label{fig:cms-pbpb-upsilons}
\end{figure}

The CMS experiment has measured the \ups~modification in \pbpb collisions at \sqrtsn = 2.76 GeV~\cite{Chatrchyan:2012lxa}. The \raa values are shown in
Figure~\ref{fig:cms-pbpb-upsilons} for the \ups(1S) and \ups(2S) states. The \ups(3S) state is so strongly suppressed that values were not
extracted. The fact that the (2S) and (3S) states are much more strongly suppressed than the (1S) state is suggestive of stronger color screening suppression 
for the larger, more weakly bound excited states. The data are well described by models in which an \ups~potential model is modified by color screening in an 
evolving quark gluon plasma~\cite{Emerick:2011xu, Strickland:2011aa}.  The statistical precision of the LHC data is expected to improve greatly during future planned 
running of the LHC program. 

The measurement of Upsilons at RHIC energy is very challenging because the production cross sections are small. So far, the measurements have been made
with low statistical precision. The three states are unresolved in PHENIX, while in STAR only the \ups(1S) state can be separated by line-shape fitting. Therefore
the existing RHIC measurements~\cite{Adare:2014hje, Vertesi:2015fua} do not provide strong constraints on theoretical models. This will change in the 
future with data from the new STAR MTD 
detector~\cite{Ruan:2009ug}, which was operational in the 2014 RHIC run, and future data from the proposed sPHENIX experiment at RHIC~\cite{Adare:2015kwa}. 

\end{document}